# Anomaly Detection in Road Networks Using Sliding-Window Tensor Factorization

Ming Xu, Jianping Wu, Haohan Wang, and Mengxin Cao

*Abstract*—Anomaly detection in road networks is vital for traffic management and emergency response. However, existing approaches do not directly address multiple anomaly types. We propose a tensor-based spatio-temporal model for detecting multiple types of anomalies in road networks. First, we represent network traffic data as a 3rd-order tensor. Next, we acquire spatial and multi-scale temporal patterns of traffic variations via a novel, computationally efficient tensor factorization algorithm: sliding window tensor factorization. Then, from the factorization results, we can identify different anomaly types by measuring deviations from different spatial and temporal patterns. Finally, we discover path-level anomalies by formulating anomalous path inference as a linear program that solves for the best matched paths of anomalous links. We evaluate the proposed methods via both synthetic experiments and case studies based on a real-world vehicle trajectory dataset, demonstrating advantages of our approach over baselines.

## I. Introduction

THE sensing of traffic situations in road networks is vital to transportation operators, and in particular, anomalies (e.g. accidents, special events) may produce rapidly diffusing traffic congestions. Road network anomaly detection can support operators in making better-informed emergency response decisions. And with a historical archive of traffic conditions, anomaly detection technologies can produce a corresponding history of anomalous events that can aid in transportation system planning as well as additional traffic analyses. Although there exists a lot of methods based on statistical theory and data mining for anomaly detection, none takes into account the variety of anomalies in road network traffic. With respect to spatio-temporal data, road network traffic data exhibits distinguishing temporal (short- and long-term) and spatial properties [1-3].

- Spatial properties. Proximal road segments are likely to have similar traffic patterns. Likewise, similar traffic trends manifest in functionally similar paths, i.e., those with functionally similar origins and destinations (e.g. homes and workplaces);
- Short-term properties. Temporally local traffic conditions are usually highly correlated, despite occasional large fluctuations;
- Long-term properties. Over large intervals of time, traffic patterns tend to be regular and stable, exhibiting specific periodicity (one day or one week).

Although the above properties represent different dimensions of traffic patterns, they are interdependent and thus should be simultaneously considered. In this paper, we find that we can identify multiple types of anomalies by leveraging the combination of spatial and temporal dimensions to significantly improve anomaly detection effectiveness. In the context of road traffic, an anomaly usually refers to a dramatic deviation from expected traffic patterns -- latent regularities of traffic variations dependent on the perspective of analysis. E.g. spatial patterns are commonalities in traffic variation among most of the links. According to deviations from various spatial and temporal patterns, we list different types of anomalous links as follows.

- Anomalous link on short-term properties (ASP). A link deviates from the short-term patterns but remains consistent with the long-term patterns.
- Anomalous link on long-term and short-term properties (ALSP). A link deviates from both the long-term and short-term patterns.
- Anomalous link on long-term properties (ALP). A link deviates from the long-term patterns, but is consistent with the short-term patterns.

We provide examples to concretely illustrate the above anomaly types. Fig. 1 presents a simple road network with 4 nodes (A-D) and 5 directed edges with an operational period of 102 days. Each day is partitioned into 144 time slots with each slot representing 10 minutes. Each grid cell on the left part represents the traffic situation in the network during one time slot, and the charts on the right part show the traffic variations with an increase of time slots on each link. We assume that each edge's traffic changes follow its own stable trend until day 100. However, link (B, D) exhibits a very different short-term traffic pattern from other links -- an example of ASP. In a real-world scenario, this anomaly might manifest itself on roads near a popular tourist attraction with large traffic flows throughout the day, contrasting roads with morning or evening rush hour

Ming Xu is with the College of Artificial Intelligence, Tianjin University of Science and Technology, Tianjin, 300222, China (e-mail: xum@bupt.edu.cn).

Jianping Wu is with the Department of Civil Engineering, Tsinghua University, Beijing, 100084, China (e-mail: jianpingwu@tsinghua.edu.cn).

Haohan Wang is with the School of Computer Science, Carnegie Mellon University, Pittsburgh PA, 15213, USA (e-mail: haohanw@andrew.cmu.edu).

Mengxin Cao is with the Language Technologies Institutes, School of Computer Science, Carnegie Mellon University, Pittsburgh PA, 15213, USA (e-mail: mcao2@andrew.cmu.edu).

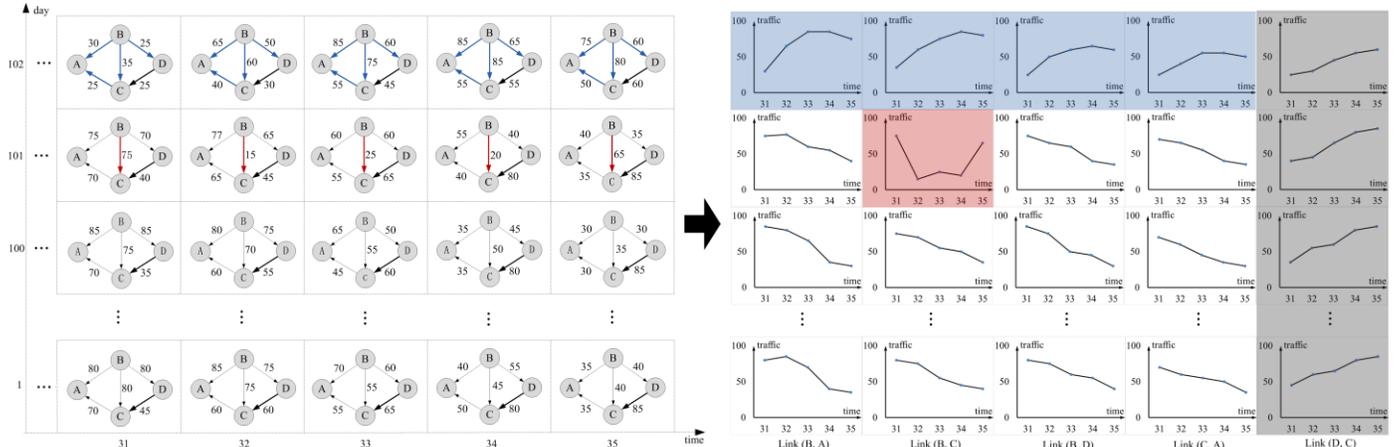

Fig. 1. A simple road network example illustrating the variety of anomaly types. Each grid cell on the left part represents the traffic situation during one time slot, and the charts on the right part show the traffic variations with an increase of time slots on each link. Each link's traffic changes follow its own stable trend until day 100. Link (D, C) (marked in black) is considered as an ASP due to its eccentric traffic variation. On day 101, link (B, C) is considered as an ALSP (marked in red), due to deviation from the short- and long-term pattern. On day 102, four links (marked in blue) are considered as ALPs, sharing similar short-term traffic pattern while deviating from their historic pattern.

intervals. On day 101, the traffic variation of link (B, C) over a few time slots is markedly different from those of other links; and moreover, it deviates from its own historical pattern. This type of anomaly may be the result of accidents or temporary traffic restrictions, and we consider it an ALSP. Observing day 102, we see a considerable number of links sharing similar short-term traffic variation while deviating from their historical pattern. A crowded event (e.g. concerts, football matches) with a large geographic area of impact may produce this type of anomaly -- which involves many ALPs. Accurate detection and identification of the above anomaly types is challenging: this requires capturing latent regularities within network traffic while simultaneously considering spatial, short-term, and long-term information. To address this, we employ a real-time, tensor-based anomaly detection method that can detect multiple anomaly types. We represent network traffic as a 3rd-order tensor and simultaneously compute spatial, short-term, and long-term patterns of traffic via tensor factorization. We then identify multiple anomaly types by measuring deviations from normal patterns on various spatial or temporal dimensions. Based on discovered link-level anomalies, we apply an optimization method to infer anomalous paths, which may provide a finer-grained understanding of anomalous events. Our contributions are as follows:

- We present a road network anomaly taxonomy based on deviations from different spatial or temporal patterns.
- We propose the sliding window tensor factorization (SWTF) algorithm to improve computational efficiency within the context of online traffic data updating.
- We improve upon the path inference method in [4] and formulate a linear program that solves for the best matched paths for detected anomalous links.
- We evaluate the effectiveness and efficiency of our proposed methods with both a synthetic experiment as well as a case study with a massive, real-world taxicab trajectory dataset.

The remainder of the paper is organized as follows. Section II provides an overview of previous work on road networks anomaly detection and tensor factorization-based anomaly detection. Section III describes the tensor representation of traffic data, SWTF algorithm for anomaly detection, and the proposed method for anomalous path inference. We present our experiments and findings in Section IV. Section V concludes our paper.

## II. RELATED WORK

In this section, we review two relevant research fields as follows.

### A. Anomaly detection in city-wide road networks

In a recent work [13], GPS trajectory data of vehicles was used to discover traffic jams. Chawla et al. [4] used PCA to detect anomalies based on taxi trajectories and then used an optimization technology to infer the anomalous paths by solving the L1 inverse problem. Pan et al. [14] first identified anomalous events according to the routing behavior of drivers, and then mined representative terms from social media to describe the detected anomalous events. Pang et al. [15] and Wu et al. [16] adapted likelihood ratio test to rapidly detect anomalies based on GPS data. To deal with widespread data sparsity in real spatial-temporal data, Zheng et al. [17] proposed a probability-based data fusion method to detect anomalies using datasets from different domains. In [18], Zheng et al. detected flawed planning of road networks using taxi trajectories. Liu et al. [19] constructed causality trees to reveal interactions among spatial-temporal anomalies and potential flaws in the design of road networks. In addition, Xu et al. [20] discovered critical nodes in road networks using a ranking algorithm based on taxi trajectories. These critical nodes can also be considered as a special class of anomalies that would cause a dramatic reduction in the network efficiency if they were to fail. Unfortunately, none of these studies address the variety of anomalies in road networks. In contrast to the abovementioned work, we detect multiple types of anomalies by combining the various spatial and temporal aspects of traffic conditions.

TABLE I
SUMMARY OF NOTATIONS

| Symbol | Descriptions |
|---|---|
| $\mathcal{A}$ ($\mathcal{A}'$) | The (updated) traffic tensor |
| $\mathcal{A}_{(i)}$ ($\mathcal{A}'_{(i)}$) | The mode-$i$ matricization of $\mathcal{A}$ ($\mathcal{A}'$) |
| $I_i$ | The size of dimension $i$ of the traffic tensor |
| $R_i$ | The rank of dimension $i$ of tensor factorization |
| $\mathcal{G}$ | The core tensor obtained by tensor factorization |
| $W^{(i)}$ ($W'^{(i)}$) | The factor matrix of dimension $i$ of $\mathcal{A}$ ($\mathcal{A}'$) |
| $Y$ ($Y'$) | The (updated) data matrix |
| $n, c$ | The dimensionality and number of data vectors in $Y$ |
| $U$ ($U'$) | The (updated) left-singular vectors |
| $G, H, Q_U, Q_V, S_U, S_V,$ | The auxiliary matrices generated in the calculation process of Bi-SVD |
| $\delta_1 \sim \delta_5$ | The thresholds of anomaly detection |
| $l_i$ | Link $i$ |
| $p_i$ | path $i$ |
| $A$ | The link-path matrix |
| $x$ | The vector of path state |
| $b$ | The vector of link state |
| $z_i, N, \mu$ | A scalar data, the number and mean of all scalar samples |
| $r_i$ | Node $i$ of region graph |
| $e, (e_i)$ | The reconstruction-error (on mode $i$) |

### B. Anomaly detection based on tensor

Many modern applications generate large amounts of data with multiple aspects and high dimensionality, for which tensors provide a natural representation. Recently, anomaly detection technologies based on tensor factorization have become increasingly popular. Zhang et al. [21] proposed a tensor-based method to detect targets in hyperspectral imagery data with both spectral and spatial anomaly characteristics. Shi et al. [23] represented a spatial-temporal data stream generated from sensor networks as an incremental tensor and proposed an incremental tensor decomposition algorithm for online anomaly detection. To detect the events in the traffic network, Fanaee-T and Gama [22] constructed a hybrid model from a topology tensor and a flow tensor, and then used a Tucker decomposition methodology with an adjustable core size. However, these studies did not consider the variety of anomalies, while our proposed methods focus on detecting different types of anomalies in traffic networks online.

## III. METHODOLOGIES

### A. Framework

As shown in Fig. 2, the framework consists of two major parts, information extraction and anomaly detection.

**Information extraction**: The framework receives real-time trajectory data stream generated by a large number of vehicles equipped GPS devices. The original trajectory data of each vehicle consists of its geospatial coordinate readings (longitude and latitude) with sampling timestamps. First, we partition a city into disjointed, equal-sized grids based on the longitude and latitude, where a grid cell denotes a region, as presented in the upper part of Fig. 3. Based on this partition, we build a region graph in which a node represents a region and a link is formed between any two adjacent regions. In practice, we can remove a link if there has never been any vehicle passing through its region pair. Next, we match the trajectory point of each vehicle to the region graph. An example is presented in bottom of Fig. 3. When the car travels from point $A$ to $B$, the volume of four links ($r_1$, $r_2$), ($r_2$, $r_3$), ($r_3$, $r_4$) and ($r_4$, $r_5$) is respectively increased by one. Then, we compute the flows of all the network links at intervals. Utilization of the region graph instead of the original road network can alleviate the issue of data sparsity and enable our model to discover more meaningful and influential events.

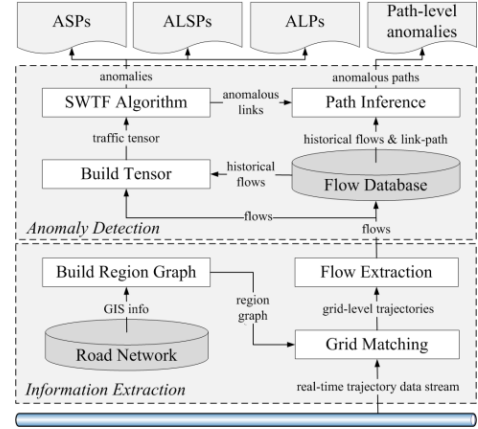

Fig. 2. Framework of the proposed method

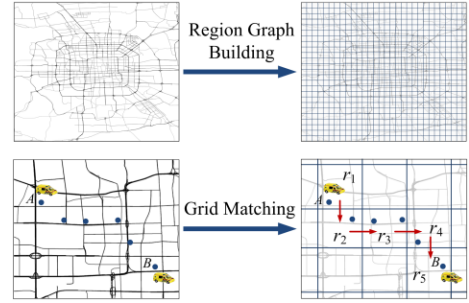

Fig. 3. Region graph building and grid matching

**Anomaly detection:** We represent the network traffic as a tensor that consists of current and historical traffic data. The key component of this framework is SWTF -- the proposed online tensor factorization algorithm for detecting multi-type anomalies. Finally, we use an optimization technology to infer the anomalous paths based on anomalous links discovered by SWTF. The path-level anomaly information is more valuable for traffic control and guidance, as it often reveals the cause of the anomaly and helps the transportation managers understand impact and scope of the events.

### B. Construction of Traffic Tensor

A tensor is a multidimensional (multiway or multimode) array or a multidimensional matrix. The order of a tensor refers to the number of dimensions. In particular, an $i$-th order ($i \geq 3$) tensor can be imagined as a hypercube of data. A scalar, vector and matrix can also be regarded as special forms of tensors, which are 0-th, first and second order tensors, respectively.

The road network traffic data can be represented with different forms of tensors. Here, we construct a 3rd-order tensor $\mathcal{A} \in \mathbb{R}^{I_1 \times I_2 \times I_3}$ with three dimensions (modes) which represent links, time slots and days, respectively, as shown in Fig. 4. $I_1$ is the total number of links in the road network; $I_3$ is the number of days. Considering that traffic data accumulates over time, we set a sliding window on the time slot dimension. $I_2$ denotes the size of the sliding window of time slot. Based on the above definitions, the indexes for the present time slot and present day are $I_2$ and $I_3$, respectively. An entry $\mathcal{A}[i_1, i_2, i_3]$ represents the traffic volume of the $i_1$-th link at the $i_2$-th time slot on the $i_3$-th day. The traffic tensor is updated as follows. At each time slot, the entries for the present time slot, $\mathcal{A}[:, I_2, :]$, denoted by blue regions, are added. $\mathcal{A}[:, I_2, :]$ consists of two parts: the entries for the present day, $\mathcal{A}[:, I_2, I_3]$, collected online and the entries for previous days, $\mathcal{A}[:, I_2, 1] \sim \mathcal{A}[:, I_2, (I_3-1)]$, obtained from the archived data. Meanwhile, the oldest entries $\mathcal{A}[:, 0, :]$, denoted by dotted line white regions, are removed from the tensor. In addition, for traffic tensor $\mathcal{A}$ shown in Fig. 4, the gray regions denote the present day entries; the solid line white regions denote the entries of previous time slots in previous days. Table I summarizes the important variables and notations in this paper.

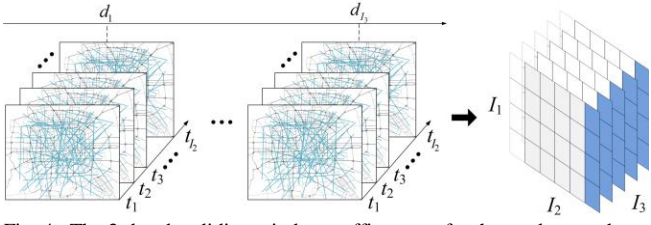

Fig. 4. The 3rd-order sliding window traffic tensor for the road network

Based on the traffic tensor, our principal concern is how to discover the implicit structures and internal relationships of traffic data. A simple method is to unfold the tensor into a large matrix along a certain dimension and then analyze it using some subspace technologies, such as PCA or NMF. However, these methods separate the relationships between different dimensions and lose some important implicit structures. For high order tensors, diagonalization and orthogonality cannot be guaranteed simultaneously. Emphasizing different aspects can obtain two different forms of decomposition: 1) CP decomposition, which preserves the diagonal form, and 2) Tucker decomposition, which emphasizes orthogonality [5]. We use Tucker decomposition for anomaly detection, since we need to find some orthogonal subspace to capture anomalies, which is described in detail in the section III. *D*. Tucker decomposition is a form of higher-order PCA and factorizes a tensor into a core tensor multiplied by an orthogonal matrix along each mode. Formally, traffic tensor $\mathcal{A}$ is factorized as

$$\mathcal{A} \approx \mathcal{G} \times_1 W^{(1)} \times_2 W^{(2)} \times_3 W^{(3)} \qquad (1)$$

where $W^{(i)} \in \mathbb{R}^{I_i \times R_i}$ $(i = 1, 2, 3)$ denotes the $i$-mode factor matrix, $R_i$ $(R_i \ll I_i)$ denotes the rank of dimension $i$; $\times_i$ denotes the $i$-mode product operator; $\mathcal{G} \in \mathbb{R}^{R_1 \times R_2 \times R_3}$ is the core tensor. The typical Tucker are HOSVD and HOOI [5]. However, they are inefficient in the context of real-time traffic data updating, for their high computational complexity.

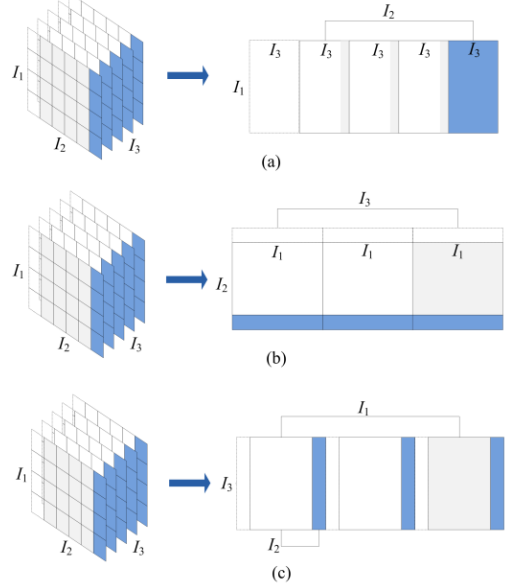

Fig. 5. Illustration of (a) mode-1, (b) mode-2 and (c) mode-3 matricization of the traffic tensor.

Therefore, SWTF is proposed to improve efficiency.

*C. Sliding-window Tensor Factorization*

Given new updated traffic tensor $\mathcal{A}'$ and old factor matrices $W^{(i)}$ $(i = 1, 2, 3)$, SWTF needs to compute the new factor matrices $W'^{(i)}$ and core tensor $\mathcal{G}'$. The computation process is summarized as 1) compute the unfolding matrix (matricization) of the tensor and its SVD for each dimension (mode); 2) output the factor matrices consisting of the principle left singular vectors. Although this process is similar to HOSVD, a challenge is how to deal with real-time streaming data and efficiently update their subspaces when new data arrives. Fig. 5 presents the mode-$i$ $(i = 1, 2, 3)$ matricizations of $\mathcal{A}'$. Note that the mode-1 and mode-3 matricizations both update multiple columns at each iteration, which results in the ineffectiveness of a class of fast subspace tracking algorithms, such as FAST [29], API [30], OPAST [31], SWASVD [28] and their variants, since such algorithms are mainly used for rank-one or rank-two updated matrices. Meanwhile, the mode-2 matricization of $\mathcal{A}'$ updates in the row vector space, which results in the subspace tracking algorithms based on incremental PCA [24, 25] or Moving Window PCA [26, 27] being inapplicable, because such algorithms only deal with the column or row space updating. To address this challenge and accurately capture the network traffic variations over time, we use the sequential bi-iteration SVD algorithm (Bi-SVD) [28].

Given a data matrix $Y \in \mathbb{R}^{n \times c} = [y_1 \ y_2 \ \cdots \ y_c]$, where $y_i$ is the $n$-dimensional data vector, SVD of $Y$ is obtained as $Y = USV^T$. When $m$ new data vectors arrive, the data matrix is updated as $Y' \in \mathbb{R}^{l \times n} = [y_m \ y_{1+m} \ \cdots \ y_{c+m}]$. Bi-SVD calculates the SVD of $Y'$ based on SVD of $Y$ with a time complexity of $O(nck)$, where $k$ is the number of dominant singular vectors. Compared with classical SVD methods with a time complexity of

## ALGORITHM I
### BI-INTERACTION SINGULAR VALUE DECOMPOSITION

**Input:** left-singular vectors $U \in \mathbb{R}^{c \times k}$, new updated matrix $Y' \in \mathbb{R}^{n \times c}$
**Output:** left-singular vectors $U'$

| | |
|---|---|
| 1 | $G \in \mathbb{R}^{l \times r} = Y'^T U$ |
| 2 | $G \xrightarrow{QR} Q_V \cdot S_V$ |
| 3 | $H \in \mathbb{R}^{n \times r} = Y' Q_V$ |
| 4 | $H \xrightarrow{QR} Q_U \cdot S_U$ |
| 5 | **return** $U' \leftarrow Q_U$ |

## ALGORITHM II
### SLIDING WINDOW TENSOR FACTORIZATION

**Input:** new updated tensor $\mathcal{A}' \in \mathbb{R}^{I_1 \times I_2 \times I_3}$, old factor matrices $W^{(i)} \in \mathbb{R}^{I_i \times R_i}$ $i = 1, 2, 3$
**Output:** new matrices $W'^{(i)}$ $i = 1, 2, 3$, new core tensor $\mathcal{G}'$

| | |
|---|---|
| 1 | For each $i \in \{1, 2, 3\}$ |
| 2 | Mode-$i$ matricization of $\mathcal{A}'$ as $\mathcal{A}'_{(i)} \in \mathbb{R}^{I_i \times \prod_{i \neq j} I_j}$ |
| 3 | $W'^{(i)} = Bi-SVD(\mathcal{A}'_{(i)}, W^{(i)})$ |
| 4 | Compute $\mathcal{G}' = \mathcal{A}' \times_1 W'^{(1)T} \times_2 W'^{(2)T} \times_3 W'^{(3)T}$ |
| 5 | **return** $W'^{(i)}$ $i = 1, 2, 3$, $\mathcal{G}'$ |

$O(\min\{nc^2, n^2 c\})$, Bi-SVD is more efficient, since $k$ is a very small number in practice. The details of Bi-SVD are summarized in Algorithm 1. In each time-step, Bi-SVD generates two auxiliary matrices $G$ and $H$. Through QR factorization of $G$ and $H$, the SVD of the updated data matrix $Y'$ can be obtained. By using Bi-SVD, SWTF efficiently computes the new factor matrices based on the previous factor matrices and new updated tensor. The details of SWTF are presented in Algorithm 2.

### D. Anomaly Detection

As discovered by previous studies [8, 32, 33] on traffic anomaly detection over the Internet, normal traffic patterns actually lie in a low dimensional subspace despite the high dimensionality of the collected traffic data. The 2-way subspace learning methods, such as PCA, can separate the low-dimensional normal traffic subspace from the abnormal traffic subspace. Through projecting onto the abnormal traffic subspace, anomalies are easily identified. Since tensor factorization is a multi-way extension of the 2-way subspace learning technology, it is natural to apply tensor factorization to anomaly detection when the data is represented as a tensor. SWTF outputs three factor matrices $W^{(1)}$, $W^{(2)}$ and $W^{(3)}$, which represent the orthogonal subspaces capturing spatial, short- and long-term traffic patterns, respectively. With respect to the $i$-th dimension $(i = 1, 2, 3)$, the normal traffic component can be calculated using $\bar{\mathcal{A}}^{(i)} = \mathcal{A} \times_i W^{(i)} (W^{(i)})^T$, and the abnormal component is $\tilde{\mathcal{A}}^{(i)} = \mathcal{A} \times_i [I - W^{(i)}(W^{(i)})^T]$. The different types of anomalies are characterized by traffic changes which deviate from the patterns on different dimensions. Thus, different anomalies can be identified by simultaneously measuring the projection value in abnormal subspaces of different dimensions. Specifically, an ASP is characterized by traffic changes following the long-term patterns but deviating from the short-term patterns. Formally, it simultaneously meets the conditions $\|vec(\tilde{\mathcal{A}}^{(2)}[i,:,:])\|_\infty > \delta_1$ and $\|vec(\tilde{\mathcal{A}}^{(3)}[i,:,:])\|_\infty < \delta_2$, where the operation $vec(\cdot)$ denotes the vectorization that stacks a tensor (or matrix) into a vector [5]; $\|\cdot\|_\infty$ denotes the infinite norm. An ALSP deviates from both the short-term and long-term patterns, i.e., it simultaneously meets the conditions $\|vec(\tilde{\mathcal{A}}^{(2)}[i,:,I_3])\|_\infty > \delta_3$ and $\|vec(\tilde{\mathcal{A}}^{(3)}[i,:,:])\|_\infty > \delta_4$. In general, ALPs are caused by a public event that may involves many links. Thus, it is much easier to detect it by observing the traffic data from a network-wide perspective rather than from a link perspective. In this case, we can determine whether an anomalous event occurs on day $j$ by examining the deviation from the network spatial patterns, i.e., $\|\tilde{\mathcal{A}}^{(1)}[:,:,j]\| > \delta_5$, where $\|\cdot\|$ denotes the 2-norm of a matrix. $\delta_1 \sim \delta_5$ are the threshold values and commonly determined using the Q-statistic [33, 34] for the abnormal space, which requires the singular values of abnormal space to be known. However, in Bi-SVD, only the $k$ largest singular values are calculated, and the remaining singular values of abnormal space are discarded. Therefore, we determine these thresholds in a simple way as

$$\delta = \mu + L\sqrt{\frac{1}{N}\sum_{i=1}^{N}(z_i - \mu)^2} \qquad (3)$$

where $z_i$ denotes a data sample; $\mu$ denotes the mean of all samples; $N$ denotes the total number of the samples; $L$ is a constant that has an important impact on selection of candidate anomalies. The determination of appropriate rank $(R_1 \sim R_3)$ of SWTF is also crucial to anomaly detection. Ideally, the principal singular vectors should capture the major traffic patterns but do not contain any anomaly information. If the number of principle singular vectors is too large, some anomalies may be considered as part of the traffic normal pattern and cannot be detected; if the number of principle singular vectors is too small, some normal traffic change may be mistaken for anomalies. Unfortunately, there is no reliable method to automatically determine these parameters. We use the simple reconstruction-error-based method also used in [21]. The reconstruction error on mode $i$ is defined as

$$e_i = \frac{\left\|\mathcal{A} - \mathcal{A} \times_i \left(W^{(i)}\right)^T \times_i W^{(i)}\right\|}{\|\mathcal{A}\|} \qquad (4)$$

We increase $R_i$ from 1 to the max number one by one. If the $i$-th singular vector captures the major traffic variance, its selection will cause a sharp drop in $e_i$; on the other size, when $e_i$ decreases slightly, it indicates that the $i$-th singular vector just captured some anomaly classes. Therefore, we select the knee

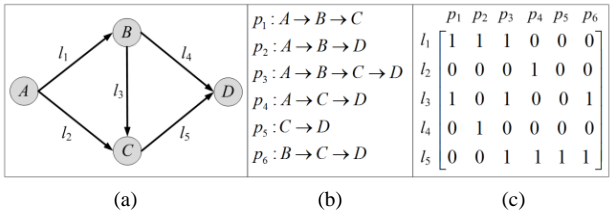

Fig. 6. A simple network with the traffic assignment. (a) Network. (b) Paths. (c) Link-path matrix

point of the $e_i$ plot (the first point of a slight decline) as the value of $R_i$.

Tensor factorization for anomaly detection have an intrinsic limitation (in analogy to PCA): normal subspace can be contaminated by anomalies from a few strongest flows, which is unavoidable for subspace method [34, 35]. However, for road networks, this limitation does not have a serious impact on the detection results. Because there can be no extremely large traffic volume on links, due to road speed limits.

*E. Inferring routes accountable for anomalies*

In this subsection, we infer the paths accountable for anomalous links detected by SWTF. Analogous studies [9, 10, 33] have emerged in the field of IP network traffic monitoring. The core idea is to formulate the anomalous path inference as a linear inverse problem and introduce the assumption of anomaly sparsity to search for the solution. The study conducted by Chawla et al. [4] is the first to apply a similar method to road network analysis. In this research, a link-path binary matrix $A \in \mathbb{R}^{m \times n}$ representing the relationship between $m$ links and $n$ paths is constructed based on vehicle trajectories. If link $i$ is on path $j$, $A_{ij} = 1$, otherwise, $A_{ij} = 0$. The binary vector $b \in \mathbb{R}^m$ denotes the states of $m$ links, which are obtained in the anomaly detection phase. If link $i$ is anomalous, $b_i = 1$, otherwise $b_i = 0$; the vector $x \in \mathbb{R}^n$ denotes the unknown states of $n$ paths to be inferred. Given these, the relationship between the states of links and paths can be represented as $Ax = b$. This is an ill-posed problem, since the paths substantially outnumber the links. This means that there are many solutions to the equation. To solve this, a constraint condition that assumes the sparsity of anomalies is introduced. Then the L1 norm minimum is used to obtain sparse solutions. In our research, we find that such a method may obtain an inexplicable negative solution, due to the high strictness of $Ax = b$. To avoid this situation, we convert this problem into a 0-1 integer linear programming problem:

$$\min \|x\|_1 \quad \text{s.t. } Ax - b \geq \mathbf{0}, x_i \in \{\mathbf{0}, \mathbf{1}\} \quad (5)$$

Such a modification guarantees to obtain a nonnegative sparse solution for path inference. However, we find that it cannot find the accurate paths related to the anomalous links in some cases. Therefore, we slightly improve on it and formulate the problem as finding the best matched paths of anomalous links:

$$\min \|Ax - b\|_1 \quad \text{s.t. } Ax - b \geq \mathbf{0}, x \geq \mathbf{0} \quad (6)$$

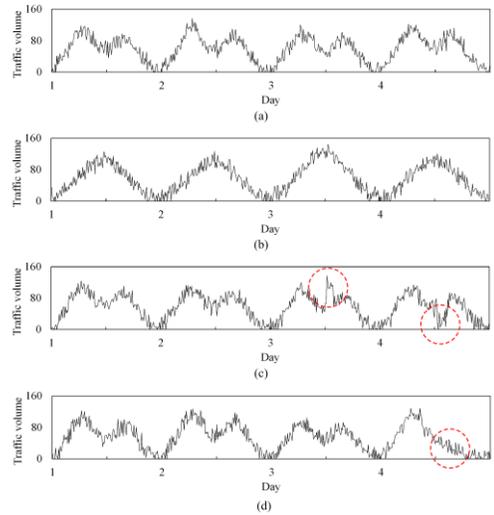

Fig. 7. Synthetic dataset containing multiple types of anomalies. (a) Normal traffic. (b) ASP. (c) ALSP. (d) ALP

Fig. 6 gives a simple network example. The topology, paths and the corresponding link-path matrix $A$ are shown in Fig. 6(a), (b) and (c). Suppose that $l_3$ is an anomalous link, i.e., $b = [0, 0, 1, 0, 0]^T$. We need to infer the unknown vector $x$ of path states. The solution of (6) is $x = [0.5, 0, 0, 0, 0, 0.5]^T$, because both $p_1$ and $p_6$ are the best matched paths of $l_3$. The solution of (5) is $x = [1, 0, 0, 0, 0, 0]^T$, which picks out one from all the paths passing through the anomalous link. If minimizing $\|x\|_1$ is subject to $Ax = b$, the solution is $x = [0, 0, 0, 0, -1, 1]^T$ which is difficult to interpret.

## IV. EXPERIMENTS

Because reliable ground truth is unavailable in real-world datasets, anomalies need to be labeled by manual inspection, which is time-consuming and prone to mistakes. Furthermore, a real dataset commonly has a very limited number of anomalies, which are insufficient to comprehensively evaluate the performance of the proposed method. Therefore, we generate synthetic datasets and inject anomalous traffic situations. Specifically, we construct the simulated networks and select a certain percentage of origin-destination (OD) node pairs for traffic assignment. We partition one day into 144 time slots, i.e., each time slot represents 10 minutes. In each time slot, the departure rate of each origin node is fixed, but it varies across different time slots, which allows us to simulate the morning and evening peaks of real traffic volume. The synthetic dataset is described as follows. Most links have the normal traffic changes with two traffic peak periods in one day, as presented in Fig. 7(a). To inject ASPs, several OD-pairs are configured with a different departure rate from the normal pattern, see (b) for an ASP example that corresponds roughly to a road near a popular tourist attraction with only one traffic peak in the daytime. When synthesizing ALSPs, we manually cut off a small number of links in some intervals, which may lead to a sharp traffic drop on these links and rise on their alternatives. An ALSP with two anomalies (marked by the red dotted circles) is presented in (c). To inject ALP events, we modify the traffic assignment for a certain day by setting a certain proportion of OD-pairs with an unusual departure rate. (d) shows the traffic change (marked by the red dotted circle) of an ALP. The dataset, which is generated from a simulated network with 3000 links operating for 600 days, contains 92 ASPs, 110 ALSPs and 45 events correlated with ALPs.

TABLE II
THE RUNNING TIME OF HOOI, HOSVD AND SWTF

| $I_1$ | $R_1$ | OTA (s) | DTA (s) | HOSVD (s) | SWTF (s) |
|---|---|---|---|---|---|
| 1000 | 10 | 32.48 | 1.77 | 0.67 | 0.37 |
| 1500 | 10 | 48.68 | 4.59 | 1.91 | 0.89 |
| 2000 | 10 | 88.25 | 10.04 | 3.27 | 1.13 |
| 2500 | 10 | 112.75 | 17.79 | 3.58 | 1.21 |
| 3000 | 10 | 172.34 | 27.68 | 5.14 | 1.37 |
| 3000 | 50 | 354.64 | 27.62 | 5.26 | 1.25 |
| 3000 | 100 | 302.74 | 27.76 | 5.31 | 1.33 |
| 3000 | 200 | 284.15 | 28.59 | 5.65 | 1.57 |
| 3000 | 300 | 401.71 | 30.39 | 6.95 | 1.88 |
| 3000 | 500 | 389.23 | 28.75 | 8.04 | 2.21 |

TABLE III
THE RUNNING TIME OF THE PROPOSED PATH REFERENCE ALGORITHM

| $I_1$ | 1000 | 1500 | 2000 | 2500 | 3000 |
|---|---|---|---|---|---|
| Running Time (s) | 2.29 | 4.25 | 9.89 | 16.42 | 21.68 |

TABLE IV
THE RECONSTRUCTION ERROR OF DTA, HOSVD AND SWTF

| $R_1$ | $R_2$ | DTA | HOSVD | SWTF |
|---|---|---|---|---|
| 1 | 1 | 0.25089421 | 0.15171678 | 0.15171678 |
| 5 | 1 | 0.25008316 | 0.15042392 | 0.15043158 |
| 10 | 1 | 0.24971717 | 0.15032091 | 0.15030403 |
| 10 | 2 | 0.15818581 | 0.14717390 | 0.14715815 |
| 10 | 3 | 0.15246301 | 0.14703289 | 0.14704949 |
| 10 | 5 | 0.14834485 | 0.14679491 | 0.14684228 |

TABLE V
PRECISION AND RECALL OF BIP AND BMP FOR ASPs

| Num. of links | Num. of paths | | Precision | | Recall | |
|---|---|---|---|---|---|---|
| | BIP | BMP | BIP | BMP | BIP | BMP |
| 7 | 4 | 4 | 0.750 | 0.750 | 0.600 | 0.600 |
| 11 | 6 | 7 | 0.667 | 0.714 | 0.667 | 0.833 |
| 18 | 8 | 9 | 0.625 | 0.778 | 0.455 | 0.636 |
| 24 | 13 | 15 | 0.692 | 0.733 | 0.563 | 0.688 |
| 26 | 14 | 17 | 0.714 | 0.706 | 0.476 | 0.571 |

TABLE VI
PRECISION AND RECALL OF BIP AND BMP FOR ALPs

| Num. of links | Num. of paths | | Precision | | Recall | |
|---|---|---|---|---|---|---|
| | BIP | BMP | BIP | BMP | BIP | BMP |
| 58 | 15 | 18 | 0.600 | 0.611 | 0.360 | 0.440 |
| 107 | 23 | 25 | 0.609 | 0.640 | 0.467 | 0.533 |
| 193 | 37 | 46 | 0.622 | 0.630 | 0.434 | 0.558 |
| 234 | 50 | 62 | 0.660 | 0.694 | 0.367 | 0.478 |
| 486 | 74 | 82 | 0.716 | 0.780 | 0.408 | 0.492 |

TABLE VII
THE NUMBER OF PATHS INFERRED BY BIP AND BMP FOR ALSPs

| Num. of links | Num. of paths | |
|---|---|---|
| | BIP | BMP |
| 9 | 4 | 5 |
| 15 | 4 | 6 |
| 18 | 7 | 11 |
| 24 | 15 | 16 |
| 33 | 17 | 20 |

*A. Efficiency*

We evaluate the efficiency of SWTF by comparing it with OTA [11], DTA [11] and HOSVD [5]. All these methods are implemented using Python and deployed on a 64-bit PC with a 2.50 GHz Intel Core i7 CPU and 8GB of RAM. We apply these algorithms to different scales of networks containing different numbers of links. Table II reports the computational time for different numbers ($I_1$) of links and different ranks ($R_1$) of the spatial factor matrix. The other parameters remain unchanged and the configurations are $I_2 = 30$, $I_3 = 200$, $R_2 = 5$ and $R_3 = 10$. Clearly, as an offline algorithm, the cost of OTA is very high and grows rapidly with the increase of $I_1$; although DTA is an incremental algorithm for online application, it requires more running time than that of HOSVD and SWTF due to the diagonalization of the large-scale covariance matrix; SWTF and HOSVD have a relatively low computational cost. By comparison, SWTF has a lower growth trend with the increase of $I_1$ or $R_1$. For the larger networks, SWTF is more than 3 times faster than HOSVD. Next, we investigate the computational time of the proposed path inference algorithm that is implemented using the CVX package in MATLAB. Table III shows the results for different scale networks. Clearly, for a road network with 3000 links, our method can find the relevant paths within half a minute, when given the anomalous links. In short, our methods are very efficient for online applications.

*B. Effectiveness*

We compare SWTF with DTA and HOSVD in tensor reconstruction error $e$ defined as

$$e = \frac{\|\mathcal{A} - \mathcal{G} \times_1 W^{(1)} \times_2 W^{(2)} \times_3 W^{(3)}\|}{\|\mathcal{A}\|} \quad (5)$$

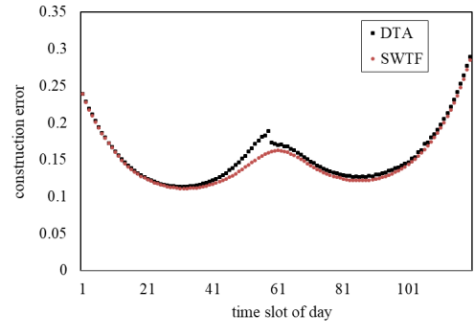

Fig. 8. The reconstruction error change over time for DTA and SWTF

All three algorithms are implemented in context of the simulated traffic data continuously arriving. The average reconstruction error of all time slots are presented in Table IV. As can be seen, SWTF and HOSVD have very similar performance, and both of them outperform DTA, especially when the ranks are small. Because DTA and SWTF are both designed for tensor stream, we also investigate the change of $e$ over time, as Fig. 8 shown. We find that $e$ decreases with the increase of network traffic volume, e.g. in the traffic peak periods in one day, which illustrates that data sparsity is an important factor affecting tensor factorization performance. Comparing with DTA, SWTF has a relative smooth variation of $e$, which implies that SWTF captures the latent traffic patterns more accurately. Fig. 9 presents the SWTF reconstructed results for a typical ASP, ALSP and ALP on the short-term and long-term dimension, respectively. For a particular anomaly type, there is a significant deviation between the original value and its reconstructed value on the dimensions associated with itself, when the anomaly occurs (marked by blue dotted circles). This also echoes our opinion introduced in the introduction.

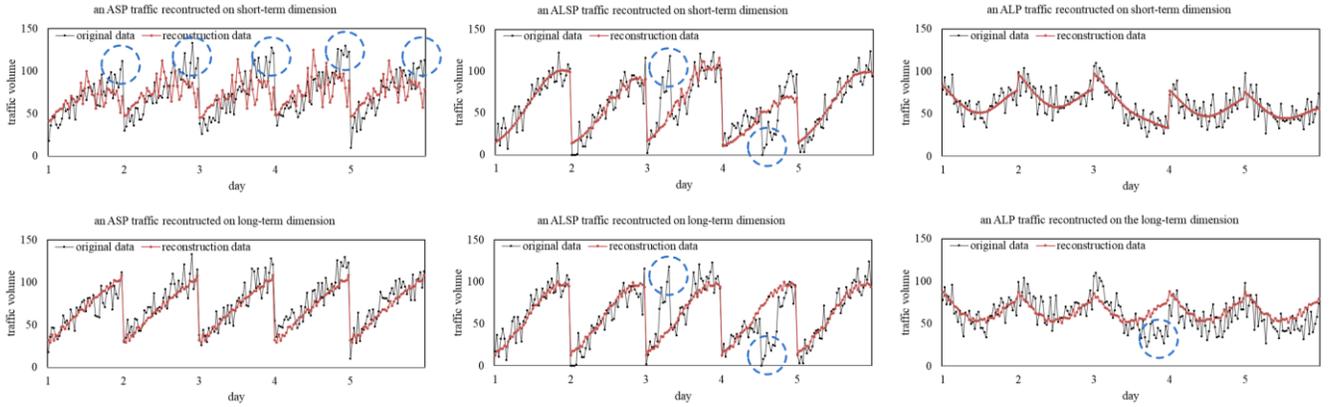

Fig. 9. SWTF reconstructed results for an ASP, ALSP and ALP on the short-term and long-term dimension

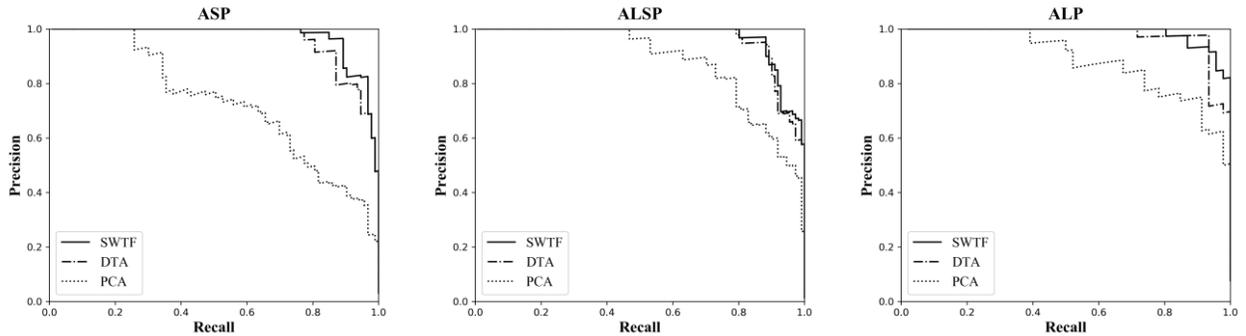

Fig. 10. The PR curves of different types of anomalies using SWTF and baseline algorithms

Since the anomaly dataset is very imbalanced, we use the precision-recall (PR) metric [13] to evaluate the effectiveness of our method. We compare SWTF with two baselines, DTA and PCA. Fig. 10 shows the PR curve of each method. The PR curve shows the tradeoff between precision and recall for different thresholds, and a large area under the curve indicates both high recall and high precision of the algorithm. In our experiments, both SWTF and DTA exhibit significant advantages over PCA, which suggests that combination of different dimensions can help distinguish the different types of anomalies and further improve performance. In addition, although SWTF and DTA have similar effectiveness, the former is more than 50 times faster than the latter. Next, we investigate the path inference algorithms based on binary integer programming (BIP) and best matched paths (BMP). In the dataset, the paths accountable for the ASPs and ALPs are labeled when injecting anomalies. Table V and Table VI present the number of inferred paths, precision and recall of BIP and BMP, respectively. As shown in the tables, BMP slightly outperforms BIP in both precision and recall. For ALSPs, it is difficult to obtain the actual paths closely related to anomalous links. Therefore, we investigate the number of inferred paths, as presented in Table VII. Compared with BIP, BMP finds slightly more anomalous paths. Based on the results, we believe that BMP can offer a better understanding of anomalous events in a short time.

### C. Real Case Studies

We conduct experiments on real trajectory data of taxicabs in Beijing to evaluate the proposed framework. The details of the datasets are as follows.

*Road network*: The road network of Beijing is segmented into 15×24 grid map, where each grid cell is a 2km×1.5km region. Based on this partition, a region graph with 360 nodes and 1362 links is formed.

*Trajectory data*: We use GPS trajectories generated by approximately 30,000 taxicabs in Beijing over a period of eight months (March 2017 ~ October 2017). The sampling interval is between 30 seconds and 60 seconds.

Fig. 11 highlights two anomalous events that are discovered by the proposed methods from the real trajectory dataset. The first event occurs during 9:30 AM-2:40 PM on 4/16/2017, as shown in Fig. 11(a). SWTF identifies this anomaly as an ALP event, since this day deviates from the spatial patterns from the network-wide prospective. If observing from the link perspective, we find that many links located in the northwest of the 2nd and 3rd ring roads share similar short-term variations but deviate their long-term patterns. By investigating the events reported by the Beijing Transportation Bureau, we find that this anomaly is associated with the Beijing International Marathon Race. The race had more than 20,000 participants and passed through some main roads such as the west ring roads etc., many of which are identified as the anomalous links by SWTF. To hold the race, traffic control was enforced on the route and a large number of vehicles had to detour, which resulted in the large flow on the links. The links with increased volume and with decreased volume are respectively marked with the blue solid and blue dotted arrows. The proposed path inference algorithm finds four OD-pairs ($r_1$, $r_2$), ($r_3$, $r_4$) ($r_5$, $r_6$) and ($r_7$, $r_8$) related to the anomalous links, where ($r_3$, $r_4$) roughly coincides with the race route.

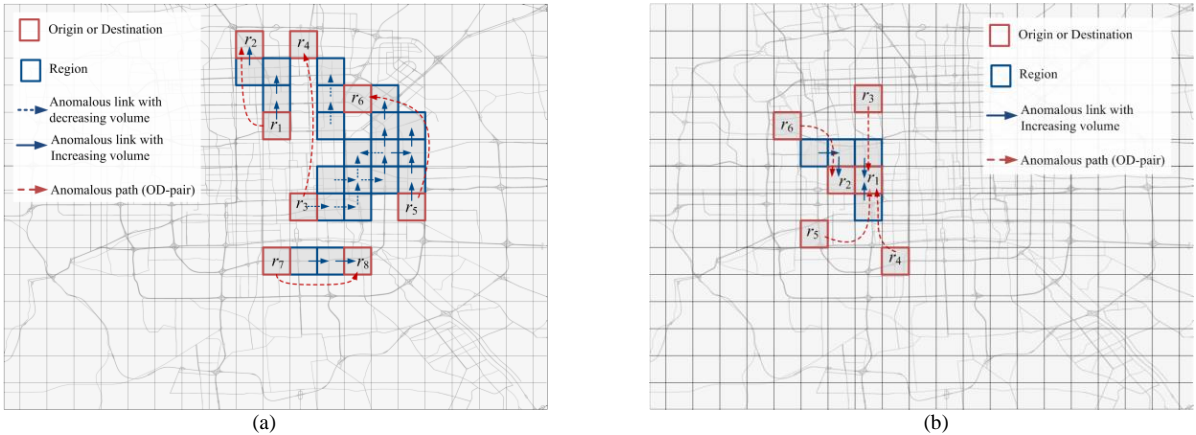

(a)          (b)

Fig. 11. The anomalies discovered by SWTF and BMP from a real trajectory dataset

Another case is presented in 11(b). The four anomalous links (marked by blue solid arrows) located near the center of the city occur at 9:50 AM 10/18/2017. SWTF identifies the four links as ALSPs, which deviate from the short- and long-term patterns. By using our path inference algorithm, we further find four anomalous OD-pairs (marked by red dotted arrows). These OD-pairs have two destination regions which are adjacent to each other. There is no record of these anomalous links in the event reports. Through Google Maps, we find that the origin regions are mainly residential areas or subway entrances, and their common destination is Beijing Children's Hospital. Moreover, these anomalous links have larger traffic volume than before. Based on these findings, we guess that this anomaly is not caused by a traffic accident but an acute infectious disease.

In addition, some anomalies that do not cause the drastic traffic change cannot be detected by SWTF, such as car crashes in the off-peak hour. To detect such anomalies, more data sources, such as videos, need to be utilized.

## V. CONCLUSIONS

In this paper, we propose a tensor-based framework to detect multiple types of anomalies in road networks using a massive real-time vehicle trajectory dataset. In this framework, we represent the network traffic as a 3rd-order tensor, and use a sliding window tensor factorization method to capture the spatial and multi-scale temporal patterns simultaneously at a low cost. Different types of anomalies can be detected by measuring the deviations from different spatial or temporal patterns. Furthermore, we infer the anomalous paths by finding the best matched paths of the discovered anomalous links. We conduct synthetic experiments and analyze a case study to evaluate the effectiveness and efficiency. The experimental results demonstrate that the proposed methods can not only detect different types of anomalies in city-wide road networks, but also provide a better understanding of the anomalies.

ACKNOWLEDGEMENT

The authors sincerely thank master student, Zhiyue Wang in the Department of Computer Science at Brown University for his help in English writing.